\documentclass[useAMS,usenatbib]{mn2e}
\usepackage{epsfig}
\usepackage{graphicx}
\usepackage{mathptmx}
\pdfoutput=1

\newcommand{\apj}{ApJ}
\newcommand{\araa}{ARA\&A}
\newcommand{\procspie}{SPIE Proc.}

\title[A UV-bright 18-min variable in NGC~1851]
{A far-ultraviolet variable with an 18-minute period in the globular cluster NGC~1851\footnote{Based on observations made with the NASA/ESA Hubble
    Space Telescope, obtained from the data archive at the Space
    Telescope Science Institute, which is operated by the Association
    of Universities for Research in Astronomy, Inc., under NASA
    contract NAS 5-26555. Some of these observations are associated
    with program GO-10184.} }
\author[D. R. Zurek, et al]{D. R. Zurek$^{1,2}$\thanks{E-mail: dzurek@amnh.org}, C. Knigge$^{1}$, T. J. Maccarone$^{3}$, D. Pooley$^{4,5}$, A. Dieball$^{6}$, 
\newauthor K. S. Long$^{7}$,  M. Shara$^{2}$, \& A. Sarajedini$^{8}$\\
$^{1}$ University of Southampton, Southampton, UK, SO17 1BJ\\
$^{2}$ American Museum of Natural History, Department of Astrophysics, Central Park West at 79th St., New York, NY, 10024\\
$^{3}$ Department of Physics, Texas Tech University, Box 40151, Lubbock, TX 79409\\
$^{4}$ Department of Physics, Sam Houston State University, Huntsville, TX 77341 \\
$^{5}$ Eureka Scientific, 5103 Fairview Dr., Austin, TX 78731 \\
$^{6}$ Argelander Institut f\"ur Astronomic, Helmholtz Institute f\"ur Strahlen- und Kernphysik, University of Bonn, Bonn, Germany\\
$^{7}$ Space Telescope Science Institute, 3700 San Martin Dr., Baltimore, MD 21218\\
$^{8}$ Department of Astronomy, University of Florida, 211 Bryant Space Science Center, Gainsville, FL 32611}

\begin{document}

\date{Accepted; 2016 May 16. Received; 2016 March 21}

\pagerange{\pageref{firstpage}--\pageref{lastpage}} \pubyear{2012}

\maketitle

\label{firstpage}

\begin{abstract}

We present the detection of a variable star with an $18.05$
minute period in far-ultraviolet (FUV) images of the globular cluster NGC
1851 taken with the Hubble Space Telescope (HST). A candidate
optical counterpart lies on the red horizontal branch or the
asymptotic giant branch star of the cluster, but it is statistically
possible that this is a chance superposition. This interpretation is
supported by optical spectroscopt obtained with HST/STIS: the spectrum 
contains none of the strong emission lines that would be expected if
the object was a symbiotic star (i.e. a compact accretor fed by a
giant donor). We therefore consider two other possibilities for the
nature of FUV variable: (i) an intermediate polar (i.e. a compact
binary containing an accreting magnetic white dwarf), or (ii) an AM
CVn star (i.e. an interacting  double-degenerate system). In the
intermediate polar scenario, the object is expected to be an X-ray
source. However, no X-rays are detected at its location in $\simeq
65$~ksec of {\em Chandra} imaging, which limits the X-ray luminosity
to $L_X \leqslant 10^{32}$~erg~s$^{-1}$. We therefore favour the AM
CVn interpretation, but a FUV spectrum is needed to distinguish
conclusively between the two possibilities. If the object is an AM CVn
binary, it would be the first such system known in any globular
cluster.

%\ltappeq

\end{abstract}


\section{Introduction}

Globular clusters (GCs) are well known for containing interacting
binaries, such as low-mass X-ray binaries (LMXBs; Clark 1975; see
Pooley 2010 for a recent review) and cataclysmic variables (CVs;
Margon et al 1981; see Knigge 2011 for a recent review), as well as
the remnants of binary interactions and collisions, such as Helium
white dwarfs (HeWDs; Cool et al 1998), blue stragglers (BSs; Sandage 1953;
Gilliland et al 1998) and millisecond pulsars (MSPs; Lyne et al
1987). As a result of the dynamical interactions taking place in dense
star clusters, the populations of LMXBs and MSPs in GCs are
significantly  enhanced relative to the Galactic field
(Clarke 1975; Camilo  \& Rasio 2005). Some theoretical simulations
suggest that the numbers of CVs and HeWDs should also be enhanced in GCs (Di 
Stefano \& Rappaport 1994; Davies 1997), though this effect is tempered
by gravitational hardening (Shara \& Hurley 2006). Conclusive
observational evidence of an enhancement or deficit of CVs and HeWDs
has yet to be found, but the importance of dynamics in regulating the
size of the CV populations in GCs has already been established (Pooley
\& Hut 2006). 

The expected enhancement of compact interacting binaries is ultimately
due to the high encounter rates in GCs (Pooley et al 2003; Heinke et al 2003; Bahramian et al 2013). 
The encounter rate is a function of the stellar density and velocity dispersion, and, since the stellar 
density in a GC is high compared to the field, interactions occur much
more frequently. NGC 1851 has an encounter rate that is one of the ten largest among 
the Galactic GCs (Bahramian et al 2013). The large number of star-star, star-binary and
binary-binary encounters results in the tightening of binaries,
exchanges of stars into and out of binaries, and stellar collisions 
(Leigh \& Geller 2012; Geller \& Leigh 2015). It is expected that these interactions 
lead to the formation of interacting compact binaries containing accreting black holes, neutron
stars and white dwarfs. 

Multiple examples of all three types of accreting compact binaries have already been discovered in GCs
(Strader et al 2012; Chomiuk et al 2013; Maccarone et al 2007; Heinke et al 2003). However, 
all accreting WDs found to date in GCs are located in
binary systems with main sequence or sub-giant donor stars
(Knigge 2012). Theoretically, accreting WDs fed by red giants (``symbiotic
stars'') or by low-mass WD donors (AM CVn stars) are also predicted,
but no such systems have been discovered to date. Here, we report the
discovery of the first strong candidate AM CVn star in a GC. This source, 
N1851-FUV1, is almost certainly an accreting WD binary system, so the main
challenge is to determine the nature of its binary companion. We
therefore begin by summarizing the main possibilities.

Interacting binaries composed of a WD and a Roche-lobe-filling
main-sequence star are known as CVs (Warner 1995; Knigge et al. 2011). The orbital periods of these
systems typically range from $\simeq 75$ minutes to about a day. The WDs in CVs can 
have varying magnetic field strengths. The magnetically weakest (``non-magnetic'') CVs have
accretion disks which extend all the way to the surface of the WD. Conversely, the
magnetically strongest CVs, known as polars, have no accretion disks at
all. In polars, the gas from the Roche-lobe-filling companion instead
flows along the magnetic field lines to impact on the magnetic
poles of the WD. Finally, CVs hosting WDs with intermedate 
magnetic fields, known as intermediate polars (IPs), contain truncated
accretion disks whose inner edge lies well outside the WD radius (Patterson 1994). Material
reaching the truncation radius then flows along the magnetic field
lines and again lands on the WD's magnetic poles. The impact region is
hotter than the surrounding WD surface, and the magnetic poles are
often not aligned with the rotational axis. As a result, the impact
regions rotate in and out of view for many lines of sight, causing
periodic variation in observed light curves that are associated with
the WD spin (Patterson 1994).

The class of interacting binaries known as AM CVs stars are similar to
CVs in that they contain a Roche-lobe-filling secondary that supplies
material to an accretion disk around a WD primary. However, while the
secondaries in CVs are typically main-sequence stars, those in AM CVns
are degenerate or semi-degenerate objects, such as Helium WDs
(HeWDs). Thus the evolutionary paths of AM CVn binaries are very
different from that of the general CV population (Solheim 2010). 
AM CVn stars are extremely compact binaries, with orbital periods less than 
$\simeq 65$ minutes (Solheim 2010). They are also relatively rare ($\simeq 40$ 
are known to date; Levitan et al. 2015), and, without dynamical enhancement, 
only about one AM CVn would be expected to exist in each GC (Nelemans et al 2001). 
In addition, AM CVns are generally faint at optical wavelengths, and the detection 
of the orbital period can be difficult.

%The class of interacting binaries known as AM CVn stars accrete material through Roche Lobe overflow into an accretion disk like other interacting binaries such as CVs. The secondary in an AM CVn is a degenerate or semi-degenerate object, such as a Helium WD (HeWD), indicating that the evolutionary path to an AM CVn binary is significantly different than that of the general CV population (Solheim 2010). AM CVn stars are extremely compact binaries, with orbital periods less than $\simeq 65$ minutes (Solheim 2010). They are also relatively rare ($\simeq 40$ are known to date; Levitan et al. 2015), and, without dynamical enhancement, only about one AM CVn would be expected to exist in each GC (Nelemans et al 2001). In addition, AM CVns are generally faint at optical wavelengths, and the detection of the orbital period can be difficult.

Symbiotic binaries are also generally accreting WD binary systems
(although a rare subset may have neutron star or black hole
primaries). However, the donor star in symbiotics is
on either the red giant branch (RGB) or the asymptotic giant branch
(AGB) (kenyon 1990; Mikolajewska 2002). In the field, symbiotic binaries have periods of hundreds to
thousands of days (Mikolajewska 2002). In a GC, tight binaries are expected to become
tighter and wide binaries are expected to become wider through
interactions with other stars (Heggie's Law: Heggie 1975). Very
wide  binaries will eventually become detached through these
interactions, so systems with periods greater than about 500 days
should be extremely rare (Heggie 1975). Near 
misses or fly-by encounters may also result in highly eccentric orbits
(Heggie \& Rasio 1996). An orbital period shorter than about 100 days
would then lead to mass exchange through Roche-lobe overflow or a
common envelope phase. Thus, in a GC, symbiotic binaries would likely
be limited to orbital periods in the range of $\simeq 100 - 500$ days and may have highly
eccentric orbits. Due to their long periods and the dominant
contribution of the donor star at longer wavelengths, symbiotic stars
will usually be missed by optical variability surveys. In fact, such 
systems can generally only be discovered by searches sensitive to
emission lines (Kenyon 1990; Mikolajewska 2002), blue/UV excess and/or fast variability associated
with accretion-induced flickering or the spin period of an accreting
magnetic WD.

Here, we present observations of a short-period variable star (N1851-FUV1) with a
significant blue excess in the core of the GC NGC 1851. Based on its
X-ray, photometric and spectroscopic properties, we suggest that it is
most likely an AM CVn star or perhaps an unusual X-ray weak IP. In
Section~2 we present a summary of the data sets used and our analysis
of the photometry, variability, spectra and X-ray observations. We
discuss the possible interpretations and fits to the SED in Section~3
and present our conclusions in Section~4.

%A summary of the data sets used in our analysis is
%presented in Section~2. Sections~3, 4, 5 and 6 present our analysis of
%the CMD position, variability properties, spectral energy distribution
%and X-ray observations, respectively. We discuss the possible
%interpretations of the data in Section~7 and, finally, present our
%conclusions in Section~8.  

\begin{table*}
\centering
\begin{minipage}{140mm}
\caption{Hubble Space Telescope imaging observations}
\begin{tabular}{@{}lcllcc@{}}
\hline
Proposal \# & Date of Observations & Instrument   &   Filter  & Number of Exposures & Total Exposure time (sec)\\
\hline
GO-6095 & 1995-10-05 & WFPC2 & F218W & 2 & 1600 \\
GO-6095 & 1995-10-05 & WFPC2 & F439W & 4 & 360 \\
GO-6095 & 1995-10-05 & WFPC2 & F555W & 2 & 46 \\
GO-5696 & 1996-04-10 & WFPC2 & F336W & 4 & 3600 \\
GO-5696 & 1996-04-10 & WFPC2 & F439W & 3 & 1200 \\
GO-7363 & 1999-03-24 & STIS/FUV-MAMA & F25QTZ & 8 & 11000 \\
GO-10775 & 2006-05-01 & ACS/WFC & F814W & 1 & 20 \\
GO-10775 & 2006-05-01 & ACS/WFC &ÊF606W & 1 & 20 \\
GO-10184 & 2006-08-15 & ACS/SBC & F140LP & 273 & 24390 \\
GO-11975 & 2009-05-02 & WFPC2 & F170W & 3 & 2100 \\
GO-11975 & 2009-05-02 & WFPC2 & F255W & 3 & 3700 \\ 
GO-11975 & 2009-05-02 & WFPC2 & F336W & 4 & 2440 \\
GO-11975 & 2009-05-02 & WFPC2 & F555W & 4 & 141 \\
\hline
\\
\\
\\
\end{tabular}
\end{minipage}
\end{table*}

\section{Data and Analysis}

We obtained FUV timeseries observations of the core of NGC~1851 using
ACS/SBC on HST (GO program 10184). We supplemented these observations
with archival WFPC2, ACS/WFC and STIS/FUV-MAMA images. These datasets
are listed in Table~1 and described in Sections~2.1 and 2.2.

Our initial analysis suggested a symbiotic binary or an IP 
as the most likely scenarios for N1851-FUV1. In order to test
these ideas, we obtained additional optical spectroscopy with HST and
deep X-ray imaging with {\em Chandra}. The spectroscopy was obtained
with the STIS/CCD/G430L instrument/detector/grating combination (GO
program 13394) and is discussed in Section~2.3. The archival ((ObsID
8966) and new (ObsID 15735) X-ray observations are described in
Section~2.4.

\begin{figure*}
\includegraphics[width=170mm]{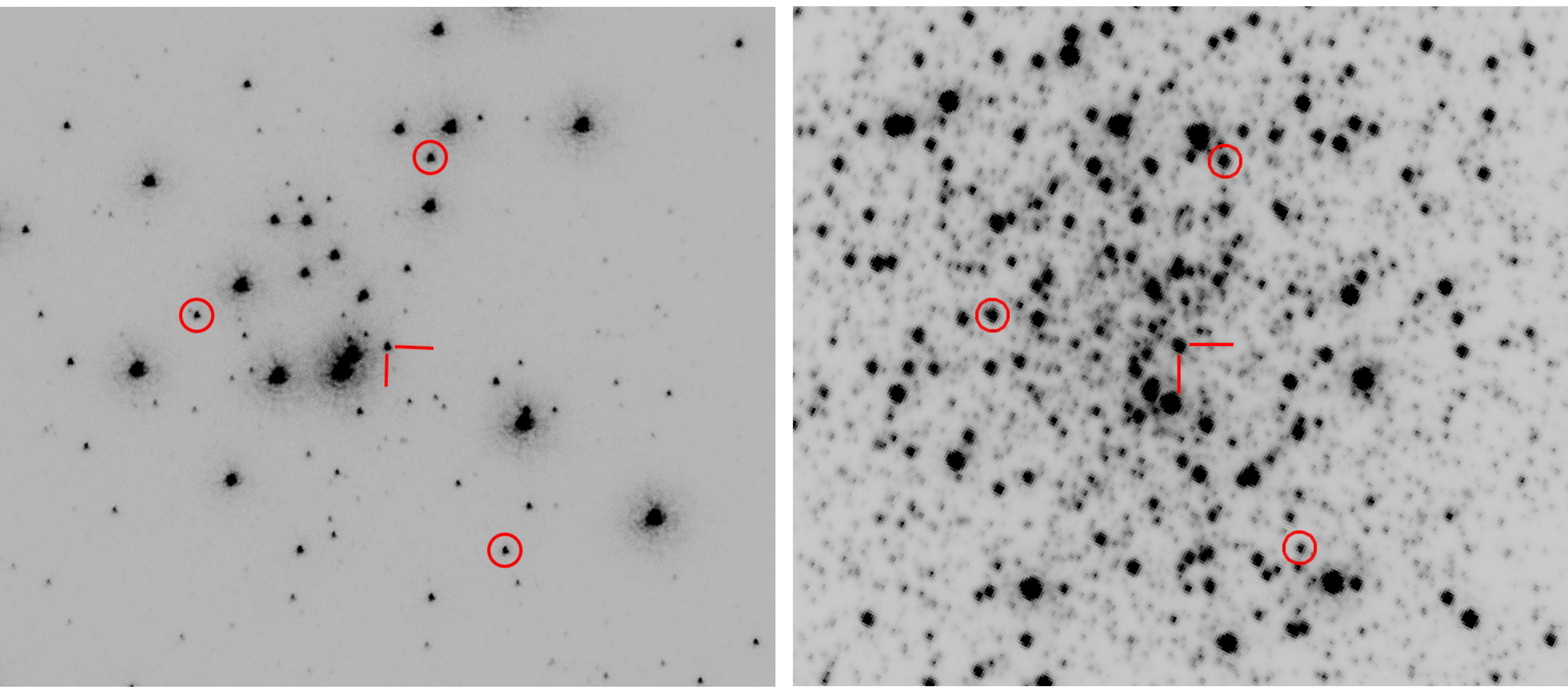}
\caption{The central 11\arcsec ~of NGC 1851 (North up and East left) with the FUV image on the left and the optical (F555W) image on the right. The position of the variable object is identified by the red tick marks. We have circled 3 stars in common between the two images to assist in identifying stars in common as the stellar density is very high in this core collapsed cluster. The field of view is 11\arcsec ~in both North-South and East-West.}
\label{fig:N1851_fig}
\end{figure*}

%\begin{figure}
%\plotone{STIS_im.eps}
%\includegraphics[width=85mm]{FUV_im.pdf}
%\caption{The FUV image of the core of NGC 1851. The position of the variable source is identified by the red tick marks. The stellar density is significantly less in the FUV as most cluster stars have very little emission at this wavelength.}
%\label{fig:fuvim}
%\end{figure}

\subsection{Optical and Ultraviolet Imaging}

\begin{figure*}
\centering
\includegraphics[width=80mm]{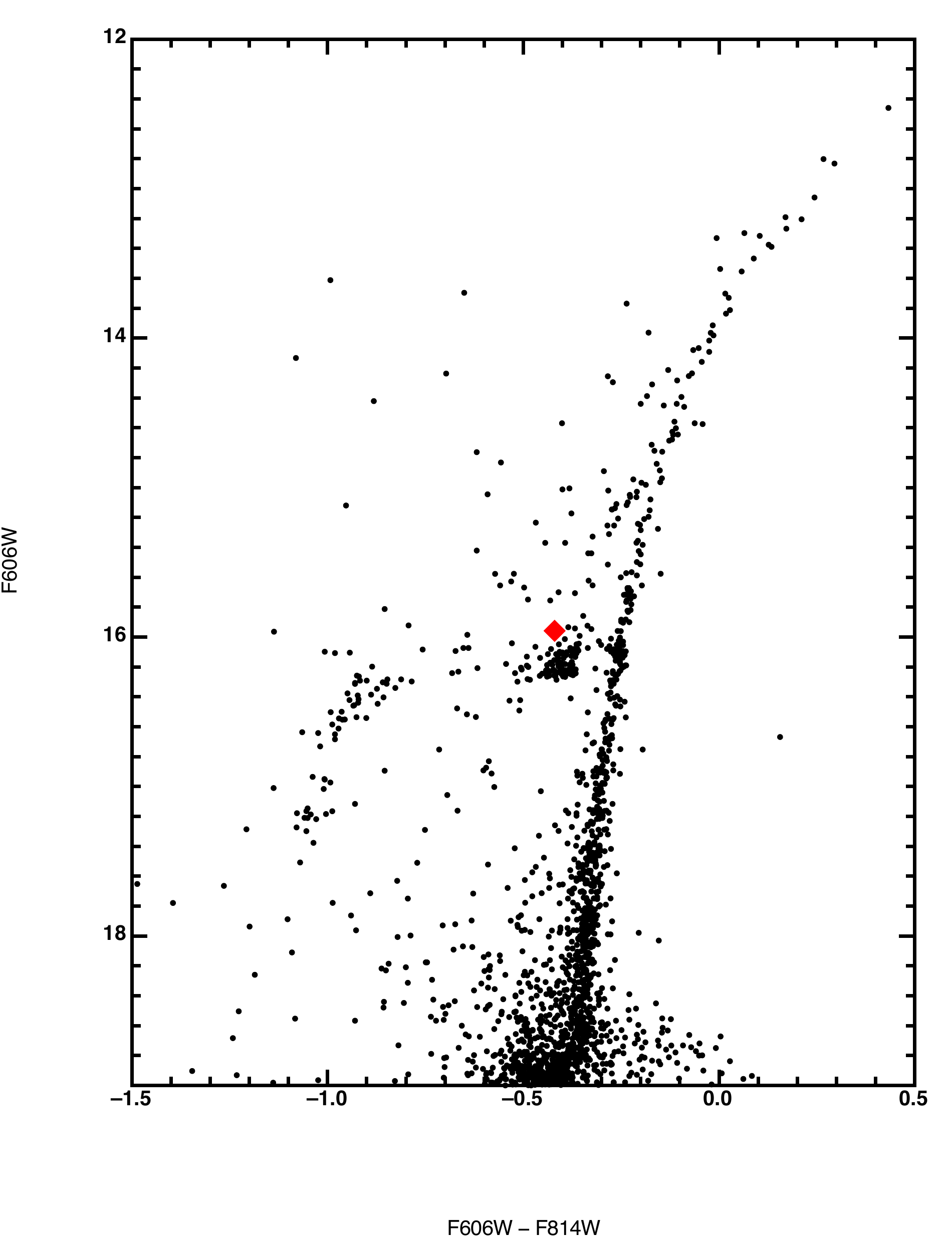}
\includegraphics[width=83mm]{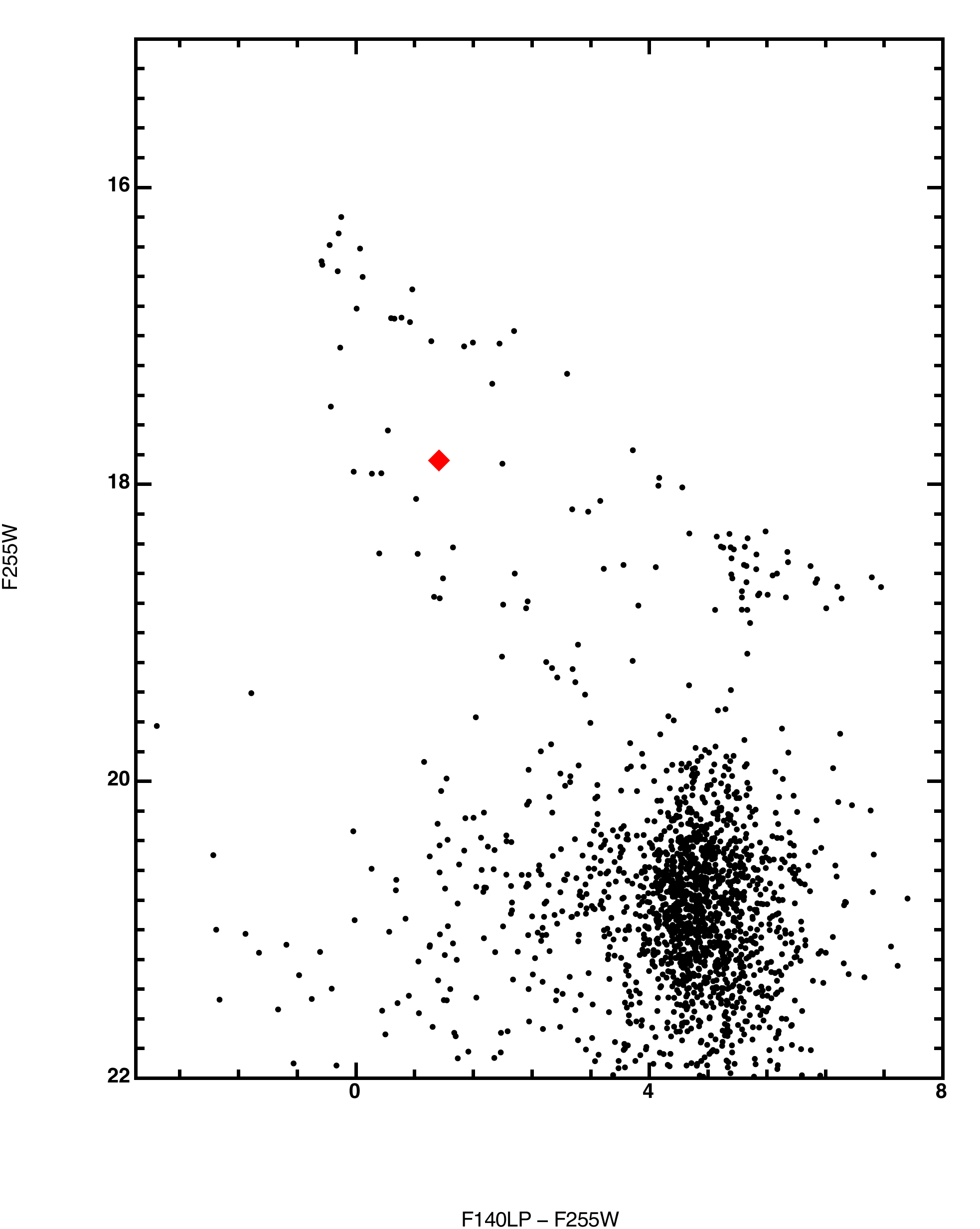}
\caption{The F606W/F814W CMD (Left Panel) from the photometric catalog of Milone et
  al. The location of the variable source is marked with a red
  diamond. It is located just above the red horizontal branch
  clump.The F255W/F140LP CMD (Right Panel) from our photometric reductions. The
  location of the variable source is marked with a red diamond. It is
  in the same locus as the bright blue straggler stars below the blue
  end of the horizontal branch.} 
\label{fig:CMD}
\end{figure*}

The WFPC2 PC images of the core of NGC~1851 contain about 50 stars per
arcsec$^2$ (Figure~\ref{fig:N1851_fig}, right panel). We therefore carried out point
spread function (PSF) fitting photometry using ALLSTAR (Stetson 94) in
order to
minimize errors due to crowding. The ultraviolet images taken with the
ACS/SBC and STIS/FUV-MAMA (Figure~\ref{fig:N1851_fig}, left panel)
require only aperture photometry, as the crowding in the ultraviolet
is considerably reduced. The ACS/WFC photometry was taken from Milone
et al (2008), who carried out PSF-fitting photometry as fully described 
in Anderson et al (2008). Our photometry
for all ultraviolet-bright sources in the cluster core will be
presented in a future paper (Zurek et al in prep).

In order to match sources across cameras, detectors and filters, we started by
considering a small set of bright sources common among the filters and
cameras. These bright sources are nearly exclusively hot horizontal
branch stars, and we used their positions as fiducial points to
calculate preliminary spatial 
transformations. These preliminary transformations were then used to
extend the matching to all detected sources and to refine the
transformations. We consider a source in one filter/camera to be
matched with a detection in another filter/camera if the separation is
at most one optical pixel or two far-UV pixels. Both optical detectors
used have pixel sizes of $\simeq$ 0.05\arcsec, while both far-UV
detectors have pixel sizes of $\simeq$ 0.025\arcsec.

Our photometric measurements for N1851-FUV1 are listed in Table~2 and
shown in the context of the relevant cluster colour-magnitude diagrams
(CMDs) in Figure \ref{fig:CMD}. In the left panel, we show the optical 
V vs V-I CMD (V=F606W \& I=F814W) constructed from the photometric
catalog of Milone et al. (2008), where V = m(F606W) and I = m(F814W). In
the right panel, we show our matched ultraviolet NUV vs FUV-NUV CMD, 
where FUV = m(F140LP) and NUV = m(F255W). In both CMDs, the location
of N1851-FUV1 is indicated with a large red diamond symbol. The key
point to note is that N1851-FUV1 appears blue in the ultraviolet CMD,
but red in the optical CMD, indicating that it must be composed of
both a hot and a cool component.
 
The crowding in the core of NGC 1851 is quite severe, so it is
important to consider the possibility that the unusual location
of N1851-FUV1 in the UV and optical CMDs is due to a chance
superposition of unrelated hot and cool objects in the cluster. We
therefore estimated the probability of such a chance superposition
happening via a simple Monte Carlo simulation based on the actual
number and locations of the relevant sources in our far-UV field of
view. Specifically, we took the positions of all far-UV sources
brighter than the main sequence (200 sources), as well as the
positions of all sources that are at least as bright and red as the
red horizontal branch in the optical CMD (205 sources). We then created
a mock data set by randomly shifting the positions of the UV and red
sources independently of each other, by amounts much greater than our
matching radius, but much smaller than the field of view. Finally, we
recorded the number of matches within this mock data set, adopting the
same matching radius as for the real data. We repeated this
simulation hundreds of times to estimate the probability of finding at
least one chance match in our data. The resulting probability is $p \simeq 8$\%.

We also looked into how often a randomly placed source matches with any object 
in the optical catalog of sources (not restricting ourselves to red giants). 
Globally the value is $\sim7$\% and within about 10 arcseconds of the cluster center 
the value is $\sim10$\%. 

These probabilities are not negligible and may even be somewhat higher near the cluster center
where N1851-FUV1 is located. Thus the possibility that the anomalous
colours of N1851-FUV1 could be due to a chance superposition must be
seriously considered. The probability that {\em one specific}
UV-bright object -- e.g. selected on the basis of exhibiting an
18-min periodic signal -- should suffer a chance superposition with a
red giant is, of course, much smaller, at $p \simeq 0.3$\%. However,
this estimate relies even more on {\em a posteriori} statistics and
should therefore be enjoyed responsibly.

\begin{figure}
\includegraphics[width=80mm]{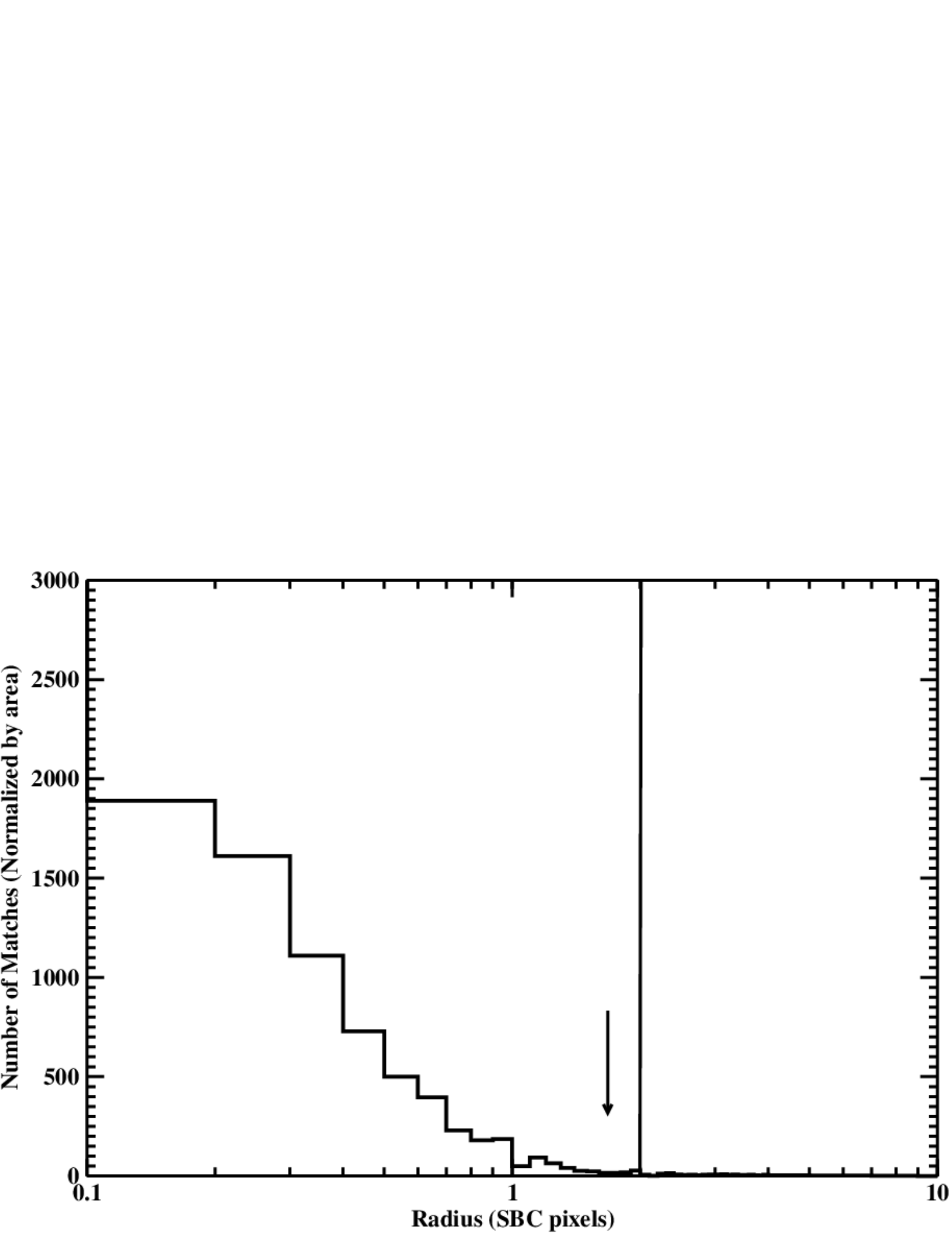}
\caption{The distribution of matches between the far-ultraviolet and the optical as a function of separation. We indicate our radius (solid vertical line) of acceptance at 2 ACS/SBC pixels. We indicate with the arrow the separation between the FUV and optical components of N1851-FUV1.}
\label{fig:radlinlog}
\end{figure}

It is worth emphasizing that our estimated chance coincidence probabilities are officially p-values: they represent the probability of finding as good a match as observed under the null hypothesis that there is no physical association(and $1-p$ is not the probability that there is an association). In order to explore this issue further, we have therefore 
examined the possibility that N1851-FUV1's optical counterpart is a chance 
superposition by examining the distribution of separations between all 
FUV sources and optical counterparts in our matched photometry 
(Figure~\ref{fig:radlinlog}). The vertical line in this figure is our 
maximum allowed separation for matches (in ACS/SBC pixels - see above). 
The arrow indicates the separation between N1851-FUV1 and its optical 
counterpart, which lies quite far in the tail of the distribution. 
We also inspected the ACS/SBC and WFC3/UVIS images in the vicinity of
N1851-FUV1 visually. By eye, there appears to be a shift of about 0.3
WFC3/UVIS pixels between the center of flux in the ACS/SBC image and
the WFC3/UVIS images. Thus, overall, there is at least circumstantial
evidence to indicate that N1851-FUV1 is, in fact, a chance superposition.

%This is not the same as the probability that there is no physical association, and $1-p$ is not the probability that there is an association. In order to explore this issue further, we have therefore examined the possibility that N1851-FUV1's optical counterpart is a chance superposition by examining the distribution of separations between all FUV sources and optical counterparts in our matched photometry (Figure~\ref{fig:radlinlog}). 
%The vertical line in this figure is our maximum allowed separation for matches (in ACS/SBC pixels - see above). The arrow indicates the separation between N1851-FUV1 and its optical counterpart, which lies quite far in the tail of the distribution.  We also inspected the ACS/SBC and WFC3/UVIS images in the vicinity of N1851-FUV1 visually. By eye, there appears to be a shift of about 0.3 WFC3/UVIS pixels between the center of flux in the ACS/SBC image and the WFC3/UVIS images. Thus, overall, there is at least circumstantial evidence to indicate that N1851-FUV1 is, in fact, a chance superposition. 

\begin{figure}
%\plotone{scargle.eps}
\includegraphics[width=80mm]{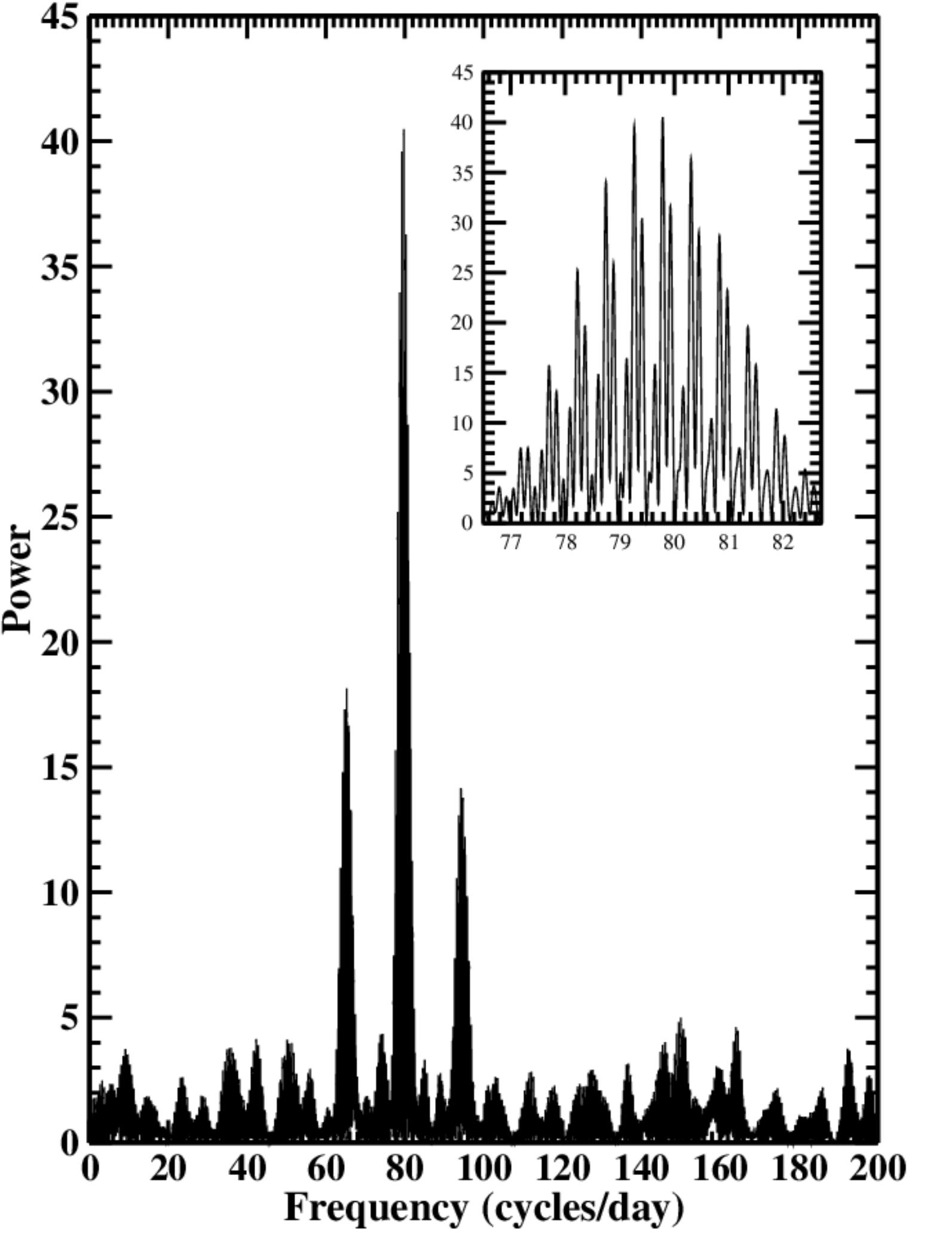}
\caption{The power spectrum of the FUV variable source. The central
  and strongest peak corresponds to $79.79$ cycles/day which is
  $18.05$ minutes. The smaller peaks on either side are the aliases of
  the stronger central peak. The inset shows a magnification of the
  strongest peak. The multiple peaks are due to aliasing. The 3 visits
  were separated by 2 days between visits 1 and 2 and 6 days between
  visits 2 and 3.} 
\label{fig:scargle}
\end{figure}

\begin{table*}
\centering
\begin{minipage}{90mm}
\caption{N1851-FUV1 magnitudes}
\begin{tabular}{@{}lcccc@{}}
\hline
%Filter & Pivot Wavelength & Year & N1851-FUV1 (ST mags) & Red Giant (ST mags) \\
Filter & Pivot Wavelength (\AA) & Year & N1851-FUV1 (ST mags) \\
\hline
F814W & 8058.8 & 2006 & 16.378 \\
F606W & 5921.6 & 2006 & 15.958 \\
F555W & 5442.9 & 1995 & 16.089 \\
F555W & 5442.9 & 2009 & 16.067 \\
F439W & 4312.1 & 1995 & 15.963 \\
F439W & 4312.1 & 1996 & 16.061 \\
F336W & 3359.5 & 1996 & 16.847 \\
F336W & 3359.5 & 2009 & 16.873 \\
F255W & 2600.4 & 2009 & 17.839 \\
F218W & 2204.4 & 1995 & 18.135 \\
F170W & 1831.2 & 2009 & 18.079 \\
F25QTZ & 1595.7 & 1999 & 18.971 \\
F140LP & 1528.0 & 2006 & 18.886 \\
\hline
\\
\\
\end{tabular}
\end{minipage}
\end{table*}

\subsection{Time-Resolved Far-Ultraviolet Photometry}

The ACS/SBC data are well suited to a search for short timescale
variations among the far-UV sources. A full discussion of all far-UV
variable sources will be presented in a future paper (Zurek et al., in
prep). There were 273 90-second exposures taken over 3 visits of 4 HST
orbits (a total of 12 orbits), spanning about 8 days from the first
visit to the third. The second visit was taken 2 days after the first
visit, with the third taking place another 6 days later. 

\begin{figure}
%\plotone{phased.eps}
\includegraphics[width=95mm]{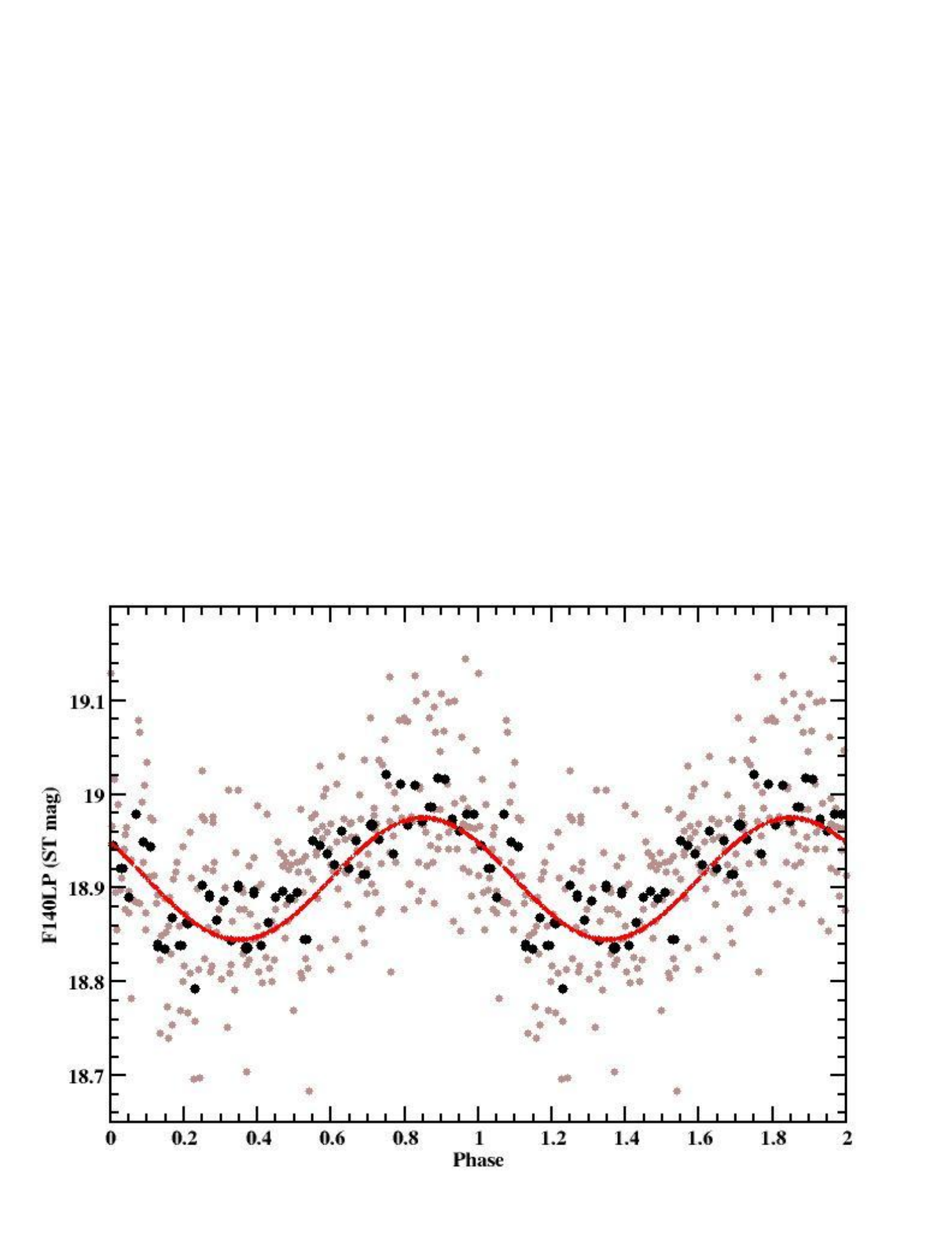}
\caption{The phased light curve of the FUV variable source. The data between the phase of 0 and 1 is duplicated between phases 1 and 2. The data has been phased around the $18.05$ minute period. All 273 individual measurements are plotted as the faint dots. The black solid dots are the average values in phase bins of $0.02$. A sine wave with an amplitude of $\sim 0.06$ mag has been fit to the data to guide the eye.}
\label{fig:phased}
\end{figure}

The fractional RMS variability for N1851-FUV1 is small ($<5\%$), and the star was not
initially classified as a variable on this basis. However, we also
carried out a search for periodic variability for all the bright
sources, by calculating the Lomb-Scargle power spectrum
for their light curves. The power spectrum for N1851-FUV1 is shown in
Figure~\ref{fig:scargle} and reveals a clear signal at $f =
79.79$~d$^{-1}$. This corresponds to a period of $P = 18.05$~min. The
far-UV light curve phased on this period is shown in
Figure~\ref{fig:phased}; for comparison, we also show a simple
sinusoid with $0.06$ mag amplitude. The precision of our period
estimate is limited to about $0.1$ minutes because of aliasing (see 
inset of Figure~\ref{fig:scargle}).  

We checked all other bright far-UV sources, but found none with the
same period. This argues against an instrumental origin of the
signal. The ultra-compact X-ray binary (4U 0513-40) has a 
17-minute period (Zurek et al. 2010), but is located $\sim6.6$
arcseconds from this source (about 150 times the
full-width-half-maximum of the PSF). 

\begin{figure}
\centering
\includegraphics[width=85mm]{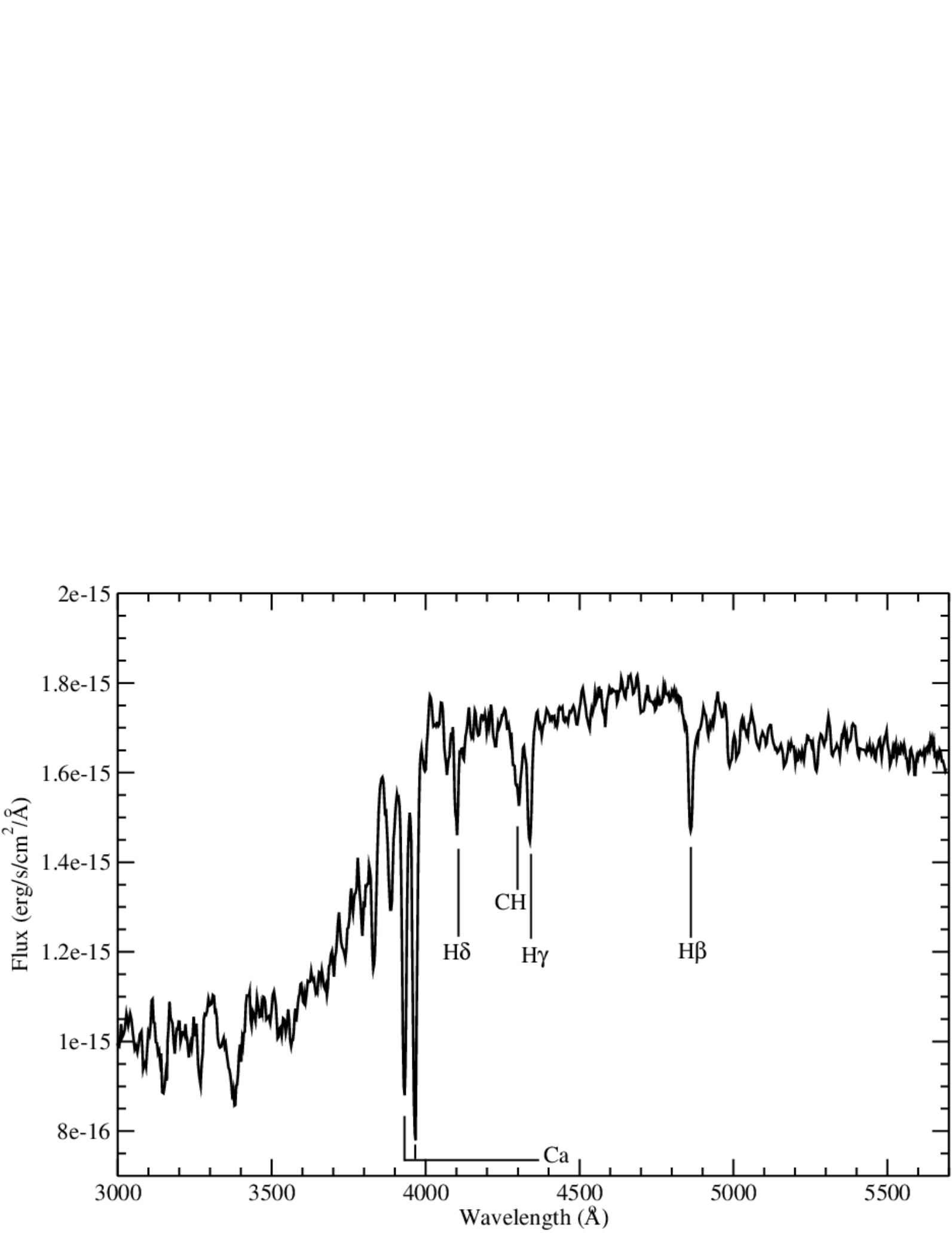}
\caption{The HST/STIS/G430L spectra of N1851-FUV1. We have indicated 
the H$\beta$, H$\gamma$, and H$\delta$ which are all in absorption. 
There is clearly no optical emission lines visible. The absorption 
feature CH also known as the g-band is a clear indication of the 
temperature as this feature is only seen in stars of spectral 
types G and K. }
\label{fig:spectra}
\end{figure}

\subsection{Optical Spectroscopy}

In order to test whether the hot and cool components contributing to
the colours of N1851-FUV1 are physically related, we obtained an
optical spectrum of N1851-FUV1 with the STIS/CCD/G430L
instrument/detector/grating combination on HST. If the system is a
close binary consisting of a hot WD and a cool giant, the giant's wind
will be irradiated by the ionizing flux of the WD. This should produce
extremely strong emission lines (such as [OIII] and the Balmer recombination
lines; Kenyon 1990), as is seen, for example, in essentially all symbiotic
binaries (Munari \& Zwitter 2002). The classification of a symbiotic binary depends on the detection of emission lines (Kenyon 1990; Mikolajewska 2002) and the most common emission lines would be within the spectral region covered in Figure~\ref{fig:spectra}. In such a symbiotic scenario, the $\simeq$ $18$~min signal would have to be associated with the spin period of an accreting magnetic WD.

However, the spectrum, shown in Figure~\ref{fig:spectra}, clearly
lacks any emission lines. If emission lines existed as in typical 
symbiotic binaries (Munari \& Zwitter 2002) we would have easily 
detected them. Instead, the spectrum is consistent with the expected 
SED of a single, cool star with $\rm T_{eff} \simeq 5200$~K. We do note that a few symbiotic binaries have been seen to enter a quiescent period where the emission lines are absent or very weak (Munari \& Zwitter 2002), indicating a change in the system. We therefore conclude that the absence of
emission lines in our data does not entirely preclude a symbiotic
nature for the system, but certainly means that we cannot classify it
as such. 

%The detection of emission lines are required to classify a system as a symbiotic binary and with the present observations N1851-FUV1 can not be classified as a symbiotic binary.

\subsection{X-ray}

The Chandra X-ray Observatory (CXO) observed NGC~1851 on three
occasions. First, on 2008 Apr 04 for an exposure time of 18.8~ks
(ObsID 8966); second, on 2015 Feb 04, for an exposure time of 19.8~ks
(ObsID 15735); and third, on 2015 Feb 07, for an exposure time of
27.7~ks. The total combined exposure time was just over 66~ks.

All data were taken in timed exposure mode with the aimpoint on the
ACIS S3 chip and telemetered in ``Very Faint'' mode. We reprocessed
all of the data with the Chandra Interactive  Analysis of Observations
(CIAO) software version 4.5 (Fruscione et al 2006) using the
latest calibration files (CALDB 4.5.5.1).  Periods of background
flaring were searched for, but none were found.  

N1851-FUV1 is located $\simeq 10$\arcsec from the bright X-ray source
4U 0513$-$40 and $\simeq$~1\farcs5 from one of the brighter of the
low-luminosity X-ray sources in the globular cluster. We therefore
applied the subpixel event repositioning (SER) algorithm
of Li et al (2004) and considered only split-pixel events,
i.e. events whose charge clouds that were spread over multiple
pixels. Such events can be located more precisely than single-pixel
events. This improves the already sharp angular resolution of Chandra
at the cost of $\simeq$~25\% of the total counts.

%\begin{figure*}
%%[!th]
%\centering
%\includegraphics[width=0.45\textwidth]{ngc1851_1.pdf}\hglue 0.04\textwidth
% \includegraphics[width=0.45\textwidth]{ngc1851_2.pdf}
%\caption{{\it Left:} A $45''\times45''$ Chandra image of the region
%  around the luminous X-ray source 4U 0513$-$40 made with 0\farcs492
%  pixels.  The location of N1851-FUV1 is marked with a red circle of
%  1\farcs2 diameter.  The locations of three low-luminosity sources
%  are marked with blue circles of the same size.  Forty-six
%  ``control'' regions of 1\farcs2 diameter at similar distances from
%  4U 0513$-$40 are shown in black. {\it Right:} A $11''\times11''$
%  image of the vicinity of  N1851-FUV1  made with 0\farcs0492 pixels
%  and using counts in the 0.5--6 keV range. } 
%\label{fig:xrayimage}
%\end{figure*}

We estimated the number of X-ray photons that may be associated
with N1851-FUV1 by counting all events in the energy range 0.5~keV -
6.0~keV within a 0\farcs5 radius 
aperture centered on its expected location. For this purpose, the
X-ray images were registered onto the optical/UV frames by a simple
boresight correction determined from 3 bright X-ray sources with
optical/UV counterparts. We then estimated the expected number of
background events in this aperture by placing identical apertures at
roughly the same distance from 4U 0513$-$40, while also avoiding other 
real X-ray sources. We find no evidence for a statistically
significant excess associated with N1851-FUV1 in any of the three
individual images, nor in the combined data. Specifically, our best
estimate for the excess number of counts in the combined data is
$6.8\pm4.6$. 

The absence of X-ray emission is primarily a constraint on the
IP scenario for N1851-FUV1. We therefore assume a
15~keV bremsstrahlung model (e.g. Patterson 1994) to convert our
measured count rate into an upper limit on the X-ray luminosity of the
system. Adopting a cluster distance of $d =
12.1$~kpc and a foreground neutral hydrogen column $\rm N_H = 4 \times
10^{20}$~cm$^{-2}$, we obtain 3-$\sigma$ upper limits on the
unabsorbed X-ray luminosity of $\rm L_{X,0.5-6}
< 5 \times 10^{31}$~erg~s$^{-1}$ (between 0.5~keV and 6~keV) or,
equivalently, $\rm L_{X,2-10} < 5 \times
10^{31}$~erg~s$^{-1}$ (between 2~keV and 10~keV). The corresponding
limit on the bolometric X-ray 
luminosity is  $\rm L_{X,bol} < 10^{32}$~erg~s$^{-1}$.

%and $L_X(14.0 - 195.0) < 5 \times
%10^{31}$~erg~s$^{-1}$. 
%The last of these -- which we provide to enable
%comparisons with the Swift/BAT sample of IPs (Pretorius \& Mukai
%2015) -- is the least certain, since it requires an extrapolation well 
%beyond the upper energy limit of the Chandra bandpass.  

%In Figure \ref{fig:xfluxcomp} we compare the X-ray to far-UV flux
%ratios of intermediate polars, AM CVns and our source. We obtained the
%far-UV fluxes from the GALEX source catalog (the FUV band of GALEX is
%similar in wavelength to F140LP) and X-ray fluxes from HEASARC source
%catalogs (XMM-Newton and/or Chandra). We find that, except for DQ Her,
%the IPs have $\rm log(F(X-ray)/F(FUV))$ ratios which are larger than
%$\sim 2.5$. 

%Figure \ref{fig:xfluxcomp} also suggest that AM CVn systems are
%distributed into two groups with the division between these two groups
%at a flux ratio of $\sim 1.7$. The AM CVn systems with a flux ratio
%less than $1.7$ have periods less than 25 minutes and conversly those
%with larger ratios have periods longer than 25 minutes.  

%Given the X-ray flux detection limit determined above we find that the
%flux ratio for N1851-FUV1 is significantly less than that of the IPs,
%however, still consistent with the AM CVn systems with periods shorter
%than 25 minutes. 

\section{Discussion}

Our analysis of all of the available observations of N1851-FUV1 has
led to the following set of constraints on the system:

\begin{itemize}
\item N1851-FUV1 is a bright far-UV source in the cluster core 
that exhibits periodic variability with $\rm P = 18.05$~min and an
amplitude of $\simeq 6\%$; 
\item its position matches that of a star on the red or asymptotic
  giant branch in the cluster, but there is a non-negligible
  possibility that this match is the result of a chance 
  superposition of two unrelated objects; 
\item its optical spectrum contains no detectable emission lines; 
\item its X-ray luminosity is $\rm L_{x,2-10} < 5 \times 10^{31}$ erg/s ($3\sigma$).

\end{itemize}

There are only a few classes of UV-bright stars that are capable of
producing such a fast periodic signal: ZZ Ceti stars (i.e. pulsating
WDs), pulsating subdwarf B stars (sdBs), symbiotic binaries, IPs and AM CVn (accreting
double WD) binaries. The symbiotic scenario is the only one in which
the blue far-UV and the red optical sources are physically
related. However, it is unlikely given the absence of emission lines in
the optical spectrum, which are a defining feature of symbiotic
systems (Munari \& Zwitter 2002). ZZ Ceti stars and sdBs are also ruled out
for N1851-FUV1: even a WD near the upper edge of the instability strip
(at $\rm T_{WD} \simeq 12,500$~K) would be more than 2 mag fainter than
observed and a pulsating sdB would be $\sim3$ magnitudes brighter in F140LP 
as it would be among the hottest horizontal branch stars (Heber 2009). 
This leaves only the IP and AM CVn scenarios, which we now
discuss in turn. Note that in both of these scenarios, the astrometric
match of the UV source to the red giant would probably have to
be a chance superposition. If the two objects were physically
associated, we would again expect to see optical emission lines
due to the photoionization of the giant's envelope (as in the case of
symbiotic binaries).

\subsection{N1851-FUV1 as an Intermediate Polar}

As explained in Section~1, intermediate polars are accreting WD
binary systems composed of a Roche-lobe-filling main sequence star and
a magnetic WD. In IPs, the accretion onto the WD takes place via an 
accretion disk that is truncated by the magnetic
field of the white 
dwarf. The gas from the inner edge of the accretion disk is then
channelled along field lines onto the magnetic poles of the WD,
producing X-ray and UV emission at and near the impact point. Since
the rotational and spin axes of the WD are generally not aligned, this
emission is modulated on the WD spin period. Both the period and
amplitude of the N1851-FUV1's far-UV variability can be explained
quite naturally in an IP scenario. 

\begin{figure}
\includegraphics[width=85mm]{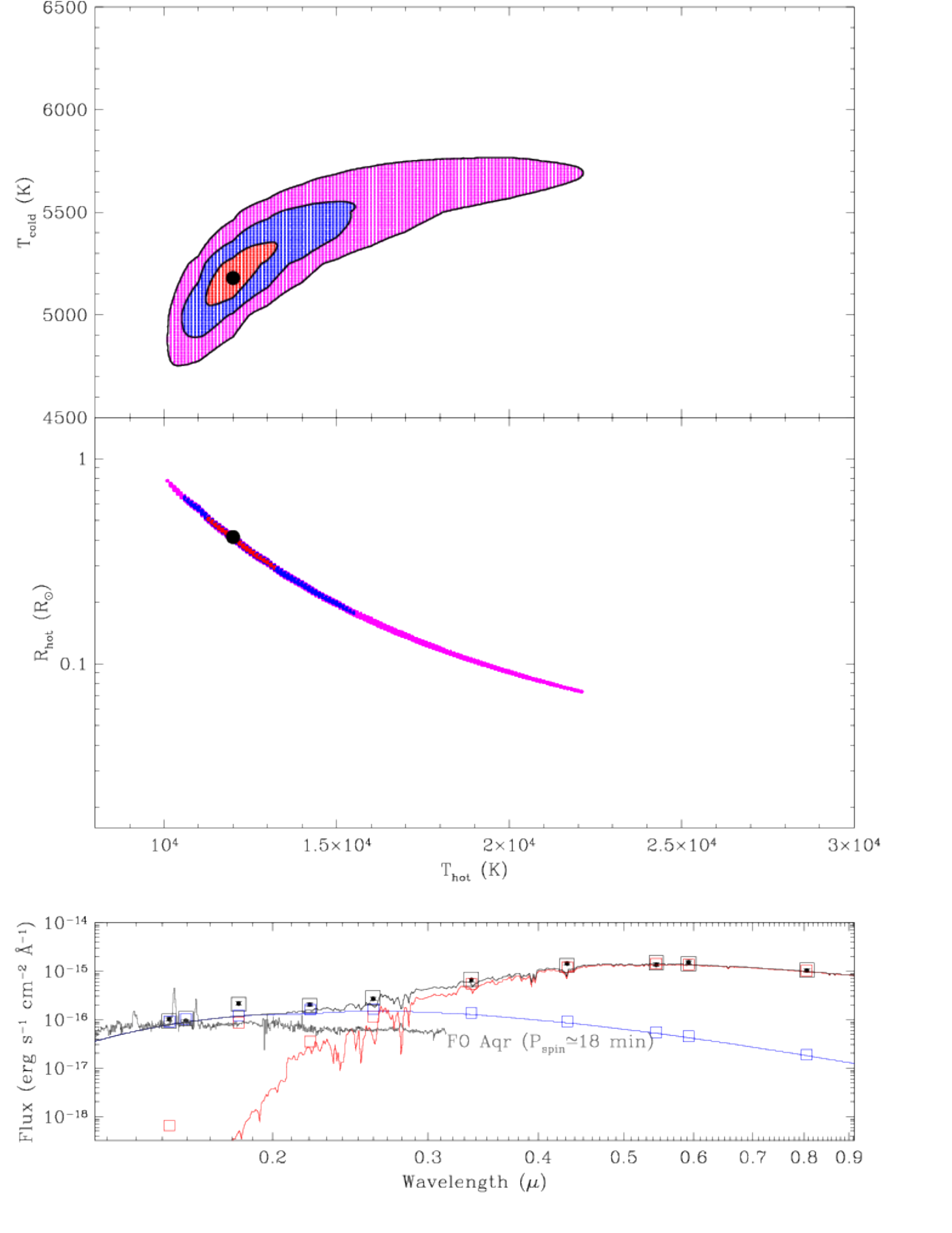}
\caption{A two-component model fit to the SED of N1851-FUV1
  representing an IP scenario. The cool
  component is modelled as a stellar model atmosphere, while the hot
  component is modelled as a blackbody (both components are modelled using
  {\sc synphot}). The top panel shows the
  best fit (black dot) and 1, 2 and 3 $\sigma$ contours in the
  parameter plane defined by the temperature of the cool component and the
accretion rate. The middle panel shows the constraints on inclination and
accretion rate. In the bottom panel, the predicted spectrum and
photometry for the best-fit cool component are shown in red, the
predicted spectrum and 
photometry for the best-fit hot component are shown in red, and the
combined best-fit spectrum and photometry
are shown in black (solid curves and open squares). The observed
photometry for N1851-FUV1 is shown by the black dots (which are mostly
located near the center of the black squares). We also show an
  example blue/ultraviolet
  spectrum of the IP (FO Aqr), which has a spin period similar to the
  period of N1851-FUV1. }  
\label{fig:fitip}
\end{figure}

Figure~\ref{fig:fitip} shows a fit to the SED of N1851-FUV1, in which
the cool component is represented by a model atmosphere with the
cluster [Fe/H] (Castelli \& Kurucz 2003), and the hot component is modelled as a blackbody. The
residuals around this fit are $\simeq 0.09$~mag, and the 
best fit parameters for the hot component are $\rm T_{hot} \simeq 12,000$~K,
$\rm R_{hot} \simeq 0.4$~R$_{\odot}$. These are reasonable numbers for the
inner accretion flow in an IP. The corresponding parameters for the
cool component are $\rm T_{cool} \simeq 5,200$~K, $\rm R_{cool} \simeq
9.5$~R$_{\odot}$. Again, these are sensible values for a red giant in
NGC~1851. We also show in Figure~\ref{fig:fitip} the UV spectrum of
FO~Aqr, a well-known field IP with a spin period of $\rm P_{spin} \simeq
18$~min, scaled to the distance and reddening of the cluster. This
spectrum lies reasonably close to the UV photometry of N1851-FUV1. We
conclude that an IP scenario can plausibly explain the UV spectrum and
variability of N1851-FUV1. 

\begin{figure}
\rotatebox{270}{\includegraphics[width=65mm]{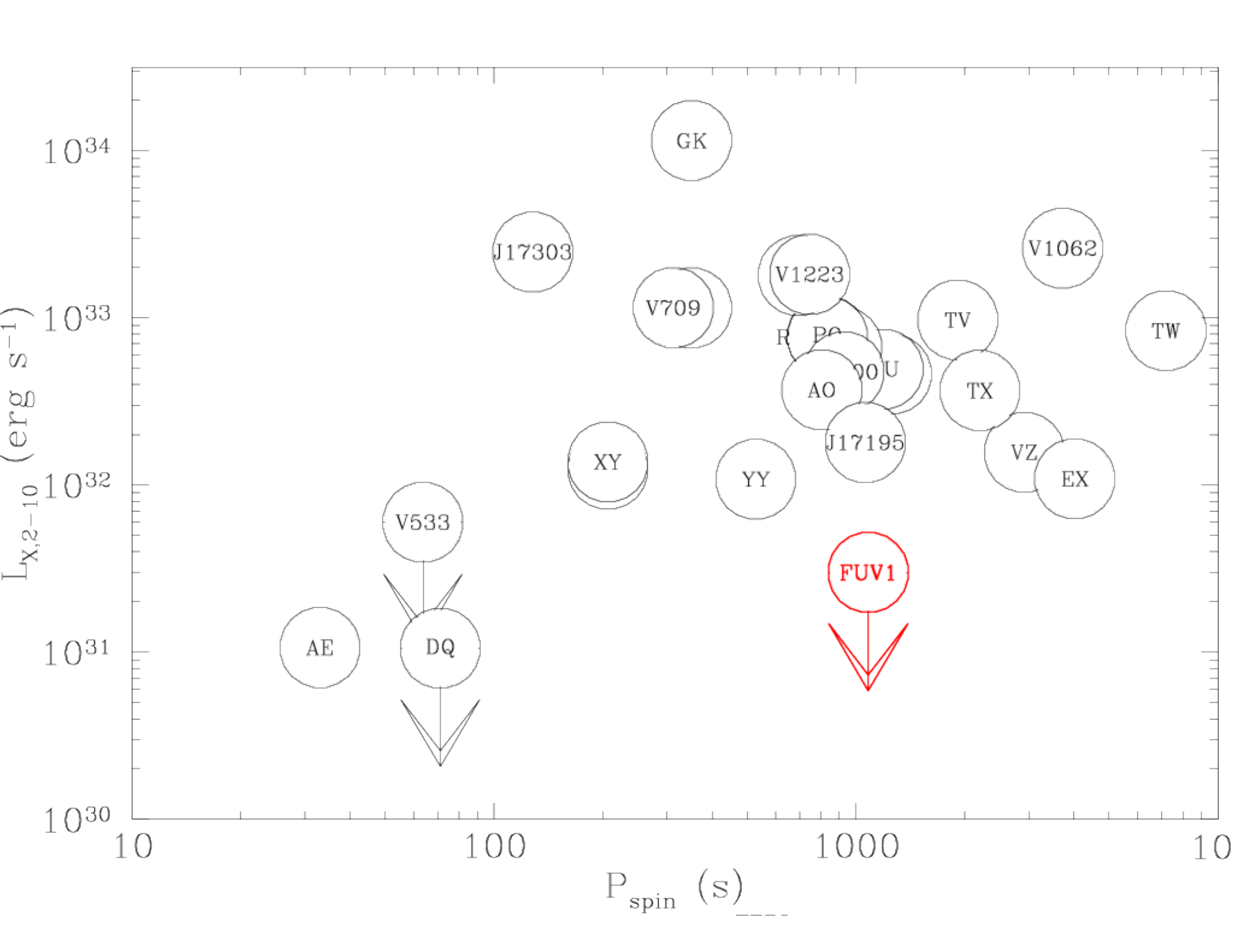}}
\caption{X-ray luminosity vs white dwarf spin period for the known
  intermediate polars (Yausa et al 2010; Patterson 1994; Pretorius \& Mukai 2014). The X-ray luminosity limit and observed period for
  N1851-FUV1 are shown in red. It is clearly much less luminous than
  sources with a comparable period. Sources with a comparable X-ray
  flux to N1851-FUV1 all have spin periods more than an order of
  magnitude smaller. The labels in the symbols identify the objects shown (e.g. AE = AE Aqr).}
%\caption{The X-ray/far-UV flux ratios of intermediate polars (black
%  histogram) and AM CVns (red histogram). The NGC 1851 source from
%  this paper is plotted as a black dot with an arrow indicating that
%  the X-ray/far-UV flux ratio is an upper limit as no X-ray flux has
%  been measured. We also indicate the $1\sigma$ and $3\sigma$
%  uncertainties on the X-ray detection limit. Note that DQ Her is the
%  only  intermediate polar with a ratio less than about 2.5.} 
\label{fig:xfluxip}
\end{figure}

However, our non-detection of the system in X-rays is a significant
challenge to an IP model. As discussed in Section~2.4, 
adopting a bremsstrahlung X-ray spectrum typical of IPs, this
non-detection implies an upper limit of $\rm L_{X,2-10} < 5 \times
10^{31}$~erg~s$^{-1}$. 
Such a low level of X-ray luminosity is unusual for IPs. This is
illustrated in Figure~\ref{fig:xfluxip}, which compares the period and
X-ray luminosity of N1851-FUV to the spin periods and X-ray
luminosities of known field IPs. X-ray fluxes for field IPs were taken
from Yuasa et al. (2010), where available, and from Patterson (1994)
otherwise. Distances and spin periods were taken from Patterson
(1994), where available, and from Pretorius \& Mukai (2014 and
personal communication) otherwise. 

Only three field IPs have X-ray luminosities that may be 
as low as that of N1851-FUV1, and all of these have extremely short 
spin periods, $\rm P_{spin} < 100$ seconds. As discussed by Patterson
(1994), the X-ray 
weakness of the fastest rotators may be due to the smaller potential
difference between the inner disk edge and the WD surface, as well to
a different accretion shock geometry in these systems. Such arguments
cannot explain the X-ray weakness of N1851-FUV1, however, given its
implied spin period of $\rm P_{spin} \simeq 1080$ seconds.

\subsection{N1851-FUV1 as a double degenerate AM CVn star}

\begin{figure}
\rotatebox{270}{\includegraphics[width=65mm]{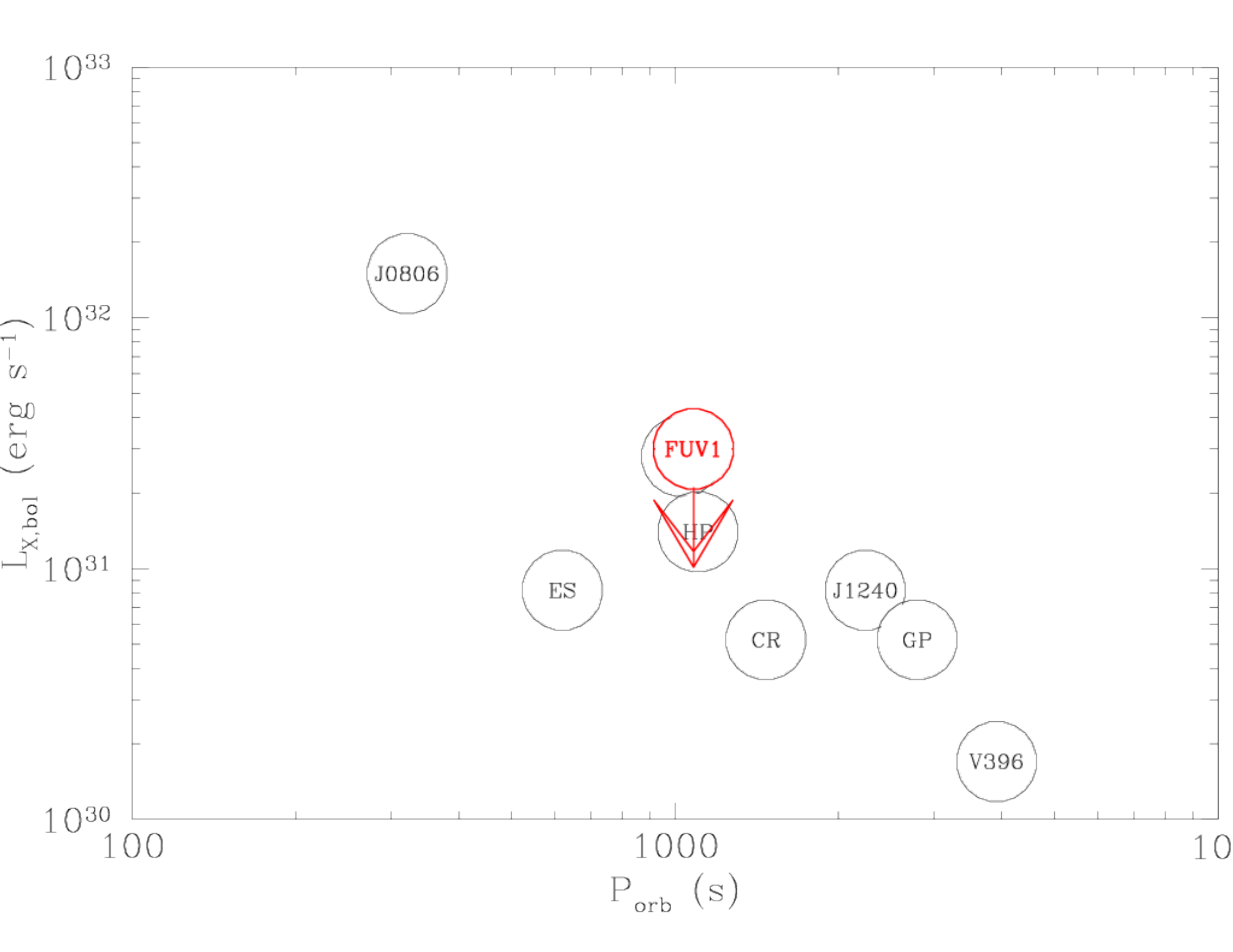}}
\caption{X-ray luminosity vs orbital period for known AM CVn binarys
  (Ramsay et al 2014). The X-ray luminosity limit and observed period for
  N1851-FUV1 are shown in red. The position of N1851-FUV1 is
  consistent with the location of known AM CVn in this diagram. The labels in the symbols identify the objects shown (e.g. ES = ES Cet).} 
\label{fig:xfluxam}
\end{figure}

AM CVn stars have orbital periods in the range 5~min - 65~min. The
period of N1851-FUV1, at $\simeq 18$ min, lies well within this
range. Moreover, in an AM CVn scenario, the lack of an X-ray detection
is actually expected, since most of these systems are X-ray weak. This
is illustrated in Figure~\ref{fig:xfluxam}, which compares the upper
limit on $\rm L_{X,bol}$ for N1851-FUV1 with the bolometric X-ray
luminosities for a sample of AM CVn stars compiled by Ramsay et
al. (2006). All but one of the systems in this sample lie below the
upper limit on N1851-FUV1.

The variability amplitude seen in N1851-FUV1 is $\approx 0.06$ magnitudes. This is about $\sim10$ times larger than that of the orbital signal seen in optical light curves of HP~Lib, an AM CVn star with a similar orbital period of $\rm P_{orb} \simeq 18$~min (Patterson et al 2002; Roelofs et al 2007). However, HP Lib also displays "superhumps", which are thought to be associated with an eccentricity of the accretion disk. The superhump period, $\rm P_{sh}$ is only a few percent longer than $\rm P_{orb}$, but the amplitude of this signal in HP Lib is much larger, at least 0.03 mag. (Patterson et al 2002). Similarly, AM CVn itself has an orbital period of $\rm P_{orb} \simeq 17.5$~min and displays optical variability amplitudes of $\simeq 0.005$ mag on $\rm P_{orb}$ and $\sim 0.01$ mag on $\rm P_{sh}$ (Skillman et al. 1999). In the slightly longer period system CP Eri ($\rm P_{orb} \simeq 28.6$~min), the superhump signal has an amplitude in excess of $\sim 0.1$ magnitudes (Armstrong et al. 2012). If N1851-FUV1 is an AM CVn star, this suggests that the 18-min signal we have detected might actually be due to superhumps, rather than to a true orbital signal. It is also worth noting, however, that the variability amplitudes tend to increase towards the blue. For example, AM CVn's variability amplitude rises to $\sim0.03$ in the FUV  (Solheim et al 1997), and multi-channel optical observations of the ultra-short period system ES Ceti  ($\rm P_{orb} \simeq 10$~min) with ULTRACM also show larger amplitudes at shorter wavelengths (Copperwheat et al. 2011). We therefore conclude that the amplitude of the 18-min signal in N1851-FUV is compatible with an AM CVn interpretation. 

%The amplitude of variation seen in N1851-FUV1 is $\approx 0.06$ magnitudes. HP~Lib with an orbital period of $\rm P_{orb} \simeq 18$~min has an amplitude $\sim10$ times smaller $\sim 0.006$ (Patterson et al 2002; Roelofs et al 2007), however, the superhump amplitude is at least 0.03 mag (Patterson et al 2002). We can not determine if the variation detected for N1851-FUV1 is due to orbital variation or a superhump. AM CVn with an orbital period of $\rm P_{orb} \simeq 17.5$~min has an amplitude of $\sim 0.012$ in the optical (Provencal et al 1995) and $\sim0.03$ in the FUV (Solheim et al 1997), and CP Eri with an orbital period of $\rm P_{orb} \simeq 28.6$~min has an amplitude of $\sim 0.1$ magnitudes (Armstrong et al. 2012). The amplitude of the photometric variation depends on the wavelength of the observations as the amplitude of the variation increases with shorter wavelength observations (see for example Figure 1 of Copperwheat et al 2011). There are only two examples, HP~Lib and AM CVn, which have a similar period to N1851-FUV1 and though the amplitude of variation is $2-3$ times larger than in these two systems N1851-FUV1 is still within the overall distribution of amplitudes among AM CVn systems.

\begin{figure}
\includegraphics[width=85mm]{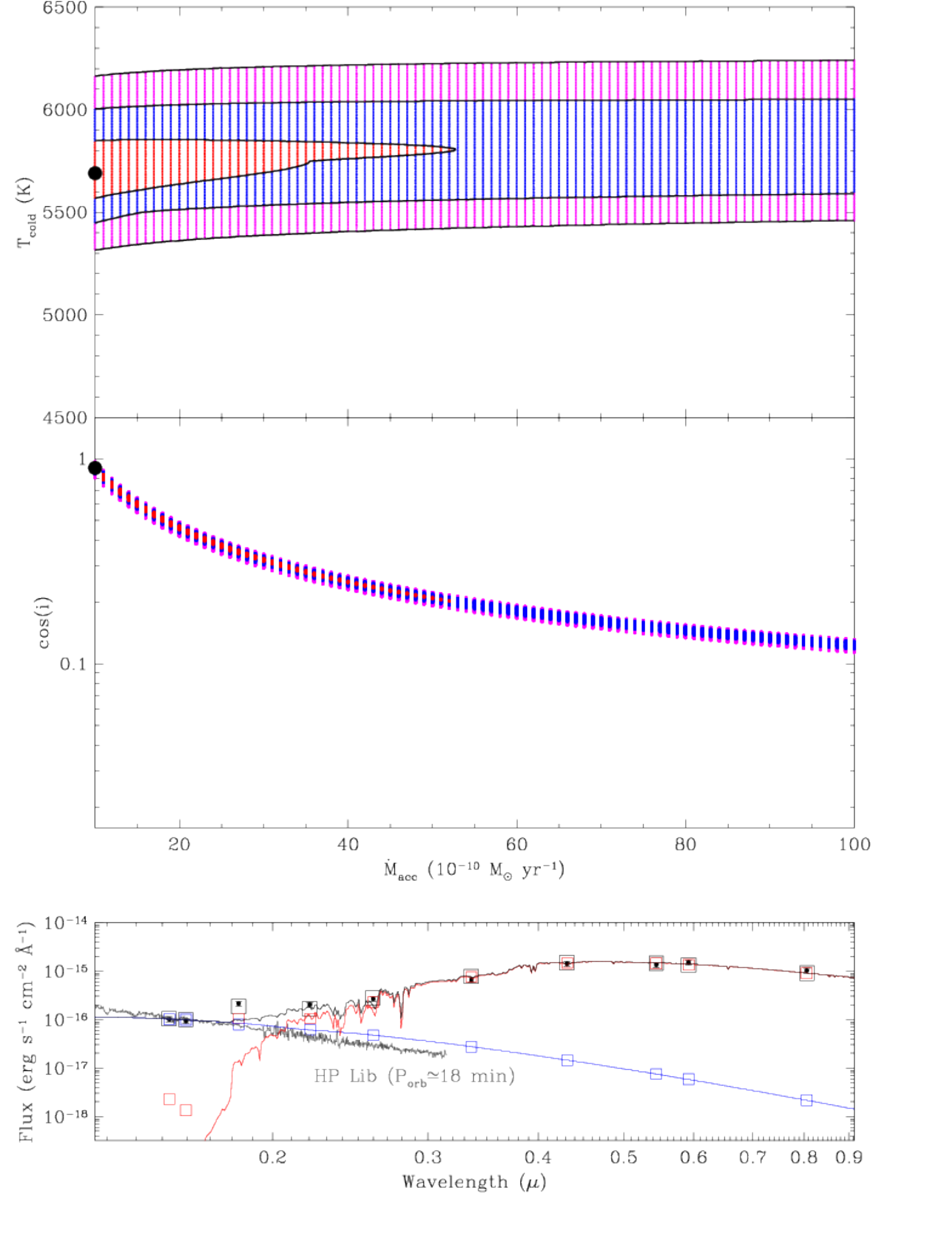}
\caption{A two-component model fit to the SED of N1851-FUV1
  representing the AM CVn scenario. The cool
  component is modelled as a stellar model atmosphere, while the hot
  component is assumed to arise from a multi-temperature accretion
  disk that locally emits as a blackbody.  The top panel shows the
  best fit (black dot) and 1, 2 and 3 $\sigma$ contours in the
  parameter plane defined by the temperature of the cool component and the
accretion rate. The middle panel shows the constraints on inclination and
accretion rate. In the bottom panel, the predicted spectrum and
photometry for the best-fit cool component are shown in red, the
predicted spectrum and 
photometry for the best-fit hot component are shown in red, and the
combined best-fit spectrum and photometry
are shown in black (solid curves and open squares). The observed
photometry for N1851-FUV1 is shown by the black dots (which are mostly
located near the center of the black squares).
We also show an example blue/ultraviolet spectrum of an AM CVn
(HP Lib) with an orbital period similar to the period of N1851-FUV1. }
\label{fig:fitdisk}
\end{figure}

The AM~CVn model is also consistent with the UV flux and the
overall SED of N1851-FUV1. This is illustrated in
Figure~\ref{fig:fitdisk}, which again shows a two-component model fit
to the SED. We once again fit the cool component with the stellar
model atmosphere spectrum appropriate for a red giant, but the hot
component is now described by a multi-temperature blackbody disk. For
the purpose of this fit, the disk is assumed to asymptotically follow
the standard $\rm T_{eff} \propto R^{-3/4}$ effective temperature
distribution, the accreting WD is assumed to have a mass of $\rm M_{1} =
0.7~$M$_{\odot}$ and a radius of $\rm R_{1} = 8 \times 10^{8}$~cm, the
mass ratio is taken to be $\rm q = M_2/M_{1} = 0.1$, and the accretion disk
is assumed to extend all the way to the last non-intersecting orbit
around the accretor, $\rm R_{d/a} = 0.6/(1+q)$ (Warner 1995). The inclination
of the disk with respect to the observer is a free parameter in this
model, subject to the constraint that $\cos{i} < 1$. The best fit
parameters for the cool component in this case are given by $\rm T_{cool}
\simeq 5700$~K and $\rm R_{cool} \simeq 7.6$~R$_{\odot}$. The best-fitting
disk parameters are $\rm \dot{M}_{acc} \simeq
10^{-9}$~M$_{\odot}$~yr$^{-1}$ and $\cos{i} \simeq 1$. With residuals
of $\simeq 0.12$~mag, formally, this fit is slightly worse than that
shown in Figure~\ref{fig:fitip}, in which the hot component was a
single temperature blackbody. However, both residuals are comparable
to the level of variability seen in the UV data. 

We also show Figure~\ref{fig:fitdisk} the UV spectrum of HP~Lib, a
well-known AM CVn star with an orbital period of $\rm P_{orb} \simeq
18$~min. Again, this lies quite close to the data in the FUV and also
to our model for the hot component. We therefore conclude that the AM
CVn model is a viable explanation for N1851-FUV1: it accounts for the
18~min periodic signal, the UV flux level and spectrum, and for the
overall SED (albeit by invoking a chance superposition with a red
giant to explain the cool component that dominates in the optical
region). Unlike the IP scenario, it also naturally accounts for the
X-ray weakness of the system. Based on all this, the AM CVn scenario
is our preferred interpretation for N1851-FUV1.

\section{Conclusion} 

We have presented the discovery N1851-FUV1, an 18~min FUV variable
star in the globular cluster NGC~1851. The position of
this hot FUV source coincides with a star on the red or 
asymptotic giant branch in the optical region. However, there is a
non-negligible chance that this astrometric match could be the result
of a chance superposition. An optical spectrum shows none of the
emission lines that might be expected if the two components were
physically associated. This makes the obvious interpretation of N1851-FUV1 as a symbiotic star in the cluster unlikely.

%This would seem to rule out the obvious interpretation of N1851-FUV1 as a symbiotic star in the cluster. 

If the cool component is unrelated to N1851-FUV1, there are two
other obvious interpretations. First, the system may be an IP, i.e. an
accreting magnetic WD that is fed by a Roche-lobe-filling main
sequence donor star. In this scenario, the 18~min periodic signal
represents the spin period of the WD. Second, the system may be an AM
CVn star, i.e. a double degenerate interacting binary in which both
the accretor and the donor are WDs. In this case, the 18~min signal is
the orbital period of the system.

Both the IP and AM CVn models can account for the observed periodic
signal, the UV flux level and the overall SED of the system. However,
no X-rays are detected from N1851-FUV1 in $\simeq$ 66~ksec of exposure
with {\em Chandra}, which implies $\rm L_{X,2-10} < 5 \times
10^{31}$~erg~s$^{-1}$. This is inconsistent with the observed X-ray
luminosities of field IPs with comparable spin periods. It is,
however, consistent with the X-ray luminosities of field AM CVn
stars. Based on this, we favour the latter scenario for N1851-FUV1,
making it the first strong AM CVn candidate known in any globular
cluster. 

In order to discriminate definitively between the IP and AM CVn
models, a FUV spectrum is required. If N1851-FUV1 is an IP, the C~{\sc iv}
and He~{\sc ii} lines will be preferentially formed in the optically thin
accretion curtain surrounding the WD and will therefore be in
emission. By contrast, the same transitions will be in absorption if
the system is an AM CVn star, since in this case the lines will be
formed in the optically thick disk. The presence or absence of a strong
Ly~$\alpha$ feature would, of course, also favour the IP or AM CVn
scenario, respectively. 

\section*{Acknowledgments}

Support for program GO-10184 and GO-13394 was provided by NASA through a grant from
Space Telescope Science Institute, which is operated by the
Association of Universities for Research in Astronomy, Inc., under
NASA contract NAS 5-26555. All of the data presented in this paper
were obtained from the Mikulski Archive for Space Telescopes
(MAST). ST ScI is operated by the Association of Universities for
Research in Astronomy, Inc., under NASA contract NAS 5-26555. We thank
Jeno Sokoloski for many helpful conversations. We are also grateful to
Retha Pretorius and Koji Mukai for helpful discussions regarding IPs
and for providing distances for some of the systems shown in
Figure~7.

%\bsp

\label{lastpage}

\end{document}